\documentclass[useAMS,usenatbib]{mn2e}
\usepackage{epsfig}

\def\simlt{\ \raise -2.truept\hbox{\rlap{\hbox{$\sim$}}\raise5.truept   %
\hbox{$<$}\ }}
\def\simgt{\ \raise -2.truept\hbox{\rlap{\hbox{$\sim$}}\raise5.truept   %
\hbox{$>$}\ }}                                                          %

\def\be{\begin{equation}}
\def\ee{\end{equation}}
\def\newline{\hfil\break}

\def\la{\mathrel{\hbox{\rlap{\hbox{\lower4pt\hbox{$\sim$}}}\hbox{$<$}}}}
\def\ga{\mathrel{\hbox{\rlap{\hbox{\lower4pt\hbox{$\sim$}}}\hbox{$>$}}}}

\title[The SZ effect from radio-galaxy lobes]{SZ effect from radio-galaxy lobes:
astrophysical and cosmological relevance}
\author[S. Colafrancesco]{S. Colafrancesco$^{1,}$ $^{2,}$ $^{3}$
 \thanks{E-mail: Sergio.Colafrancesco@mporzio.astro.it (SC)
}\\
 $^{1}$ASI Science Data Center, ASDC c/o ESRIN,
            Via G. Galilei 00044 Frascati, Italy\\
 $^{2}$Agenzia Spaziale Italiana,
            Unit\`a Osservazione dell'Universo,
            Viale Liegi 26 00198 Roma, Italy\\
 $^{3}$INAF - Osservatorio Astronomico di Roma, Via Frascati 33,
  Monteporzio 00040, Italy
 }

\begin{document}

\date{Accepted 2008 January 14.  Received 2008 January 14; in original form 2007 February 19}

\pagerange{\pageref{firstpage}--\pageref{lastpage}} \pubyear{2008}

\maketitle

\label{firstpage}

\begin{abstract}
We derive the SZ effect arising in radio-galaxy lobes that are filled with high-energy,
non-thermal electrons. We provide here quantitative estimates for SZ effect expected from
the radio galaxy lobes by normalizing it to the Inverse-Compton light, observed in the
X-ray band, as produced by the extrapolation to low energies of the radio emitting
electron spectrum in these radio lobes. We compute the spectral and spatial
characteristics of the SZ effect associated to the radio lobes of two distant radio
galaxies (3C294 and 3C432) recently observed by Chandra, and we further discuss its
detectability with the next generation microwave and sub-mm experiments with arcsec and
$\sim \mu$K sensitivity. We finally highlight the potential use of the SZE from
radio-galaxy lobes in the astrophysical and cosmological context.
\end{abstract}


\section{Introduction}

The Sunyaev-Zel'dovich effect (hereafter SZE, Sunyaev \& Zel'dovich 1972, 1980) is a
powerful probe of the physical conditions of electronic plasmas in astrophysical and
cosmological context (see Birkinshaw 1999 for a review).
A SZE can be produced by thermal (hot and/or warm) electrons in the atmospheres of
galaxies and galaxy clusters (Birkinshaw 1999, Itoh et al. 1998, Colafrancesco et al.
2003), by non-thermal (and relativistic) electrons in clusters and in the cavities
produced by AGN radio lobes (Ensslin \& Kaiser 2000, Colafrancesco et al. 2003,
Colafrancesco 2005) and by the secondary electrons produced by Dark Matter annihilation
in cosmic structures (Colafrancesco 2004), beyond the kinematic SZE related to the bulk
motion of such plasmas.\\
A SZE from the lobes of radio galaxies is also expected (Ensslin \& Kaiser 2000,
Colafrancesco et al. 2003, Blundell et al. 2006, Colafrancesco 2005, 2007) but it has not
been detected yet. Nonetheless, several indirect limits can be set to the SZE from
radio-galaxy lobes using both the available radio and X-ray information of these
structures.

In fact, detailed studies of radio-galaxy lobes (e.g., Brunetti et al. 2002, Harris \&
Krawczynski 2002, Kataoka et al. 2003, Croston et al. 2005, Blundell et al. 2006 end
references therein) have shown that these extended structures contain electrons that have
passed through the inner jets and the hotspots and are diffusing on larger-scale lobes
where they are currently available to produce inverse-Compton X-ray emission and the
relativistic SZ effect which we want to discuss here, beyond the low-frequency
synchrotron radio emission.\\
In this context, Erlund et al. (2006) found that the extended X-ray emission observed by
Chandra along the lobes of three powerful, high-redshift radio galaxies 3C 432, 3C 294
and 3C 191 is likely due to Inverse Compton scattering (ICS) of Cosmic Microwave
Background (CMB) photons by the relativistic electrons confined in the lobes. This is
consistent with previous findings of X-ray emission from the lobes of FR-II radio
galaxies (like 3C223 and 3C284, see Croston et al. 2004) attributed to the same ICS of
CMB photons.\\
Several studies have been already performed on inverse-Compton scattering in the lobes of
radio galaxies and lobe-dominated quasars at low redshift (see e.g. Hardcastle et al.
2002; Comastri et al. 2003; Croston et al. 2004; Bondi et al. 2004; Kataoka \& Stawarz
2005; Croston et al. 2005; Hardcastle \& Croston 2005).
These studies have shown that the X-ray emission from the lobes of radio galaxies is
two-sided and is not dominated by the jet inverse-Compton X-ray emission, as happens
instead in powerful, core-dominated quasars (see, e.g., the prototypical case of  PKS
0637-752,  Tavecchio et al. 2000). Only if relativistic beaming effects are important,
the jets will dominate the synchrotron and/or inverse-Compton emission on kpc scales. On
the contrary, it is clear that relativistic beaming is never important in extended radio
lobes.

Detailed studies of the ICS X-ray emission provide constraints to the energetics of the
electron population in the lobes. Specifically, a lower limit on the electron energy in
the lobe, which is of order of a few times $10^{59}$ erg, has been derived for
relativistic electrons that have an energy cutoff $\gamma_{min} \sim 10^3$ (see, e.g.,
Erlund et al. 2006), where $E= \gamma m_e c^2$ for relativistic electrons.
Such a value of the lobe energy has to be considered, however, as a lower limit since the
total energy estimate depends on the shape and on the extension of the electron spectrum
(see eqs.\ref{eq.espectrum} and \ref{eq.energy} below) and might increase by even a large
factor for lower values of $\gamma_{min}$ and steep electron spectra.

Under the assumption that the X-ray emission from radio-galaxy lobes is due to ICS of CMB
photons, a simple energy argument yields a value of the minimum electron energy $E_{min}
\approx 0.35 GeV (E_X/keV)^{1/2}$ that corresponds to $\gamma_{min} \approx 484$ in order
to be detected by Chandra in the energy range $0.5 - 3.0$ keV.\\
There are also various arguments indicating that realistic values of $\gamma_{min}$ are
likely in the range $ \sim 1 - 10^2$ (see e.g. Hardcastle et al. 2002; Comastri et al.
2003; Croston et al. 2004; Bondi et al. 2004; Kataoka \& Stawarz 2005; Croston et al.
2005; Hardcastle \& Croston 2005).
In this respect, direct constraints on the value of $\gamma_{min}$ of the order of
$\gamma_{min} \sim {\rm a \, few} \cdot 10^2 - 10^3$ can be derived in radio galaxy
hotspots (see, e.g., Hardcastle 2001).
Since we expect that there is adiabatic expansion of the relativistic plasma out of the
hotspots, the value of $\gamma_{min}$ in the radio galaxy lobes should then be
substantially lower than the previous estimate, and very likely $< 10^2$ (see, e.g., also
the discussion in Blundell et al. 2006).\\

Assuming that a unique electron spectrum is responsible for the radio (i.e. synchrotron)
and X-ray (i.e. IC) emission features of the lobes, the same electron population that
produces X-rays by ICS inevitably produces also an associated SZE which has a specific
spectral shape that depends on the shape and on the energy extention of the relative
electronic spectrum. Such SZE emission is also expected to be co-spatial with the
relative ICS X-ray emission.\\
We present here predictions for the relativistic, non-thermal SZE produced by the
electrons diffusing within powerful radio-galaxy lobes.
A relevant aspect that motivated our study is that more reliable estimates of the lobe
SZE are obtained by normalizing the energy of the electronic population residing in the
lobes to the available X-ray observations of two specific sources, 3C432 (at $z=1.785$)
and 3C294 (at $z=1.779$).
We study the amplitude and the spatial distribution of the SZE from the lobes of these
sources and provide estimates of detectability with the next generation SZE instruments
with arcsec and $\mu$K sensitivity (like ALMA and/or SPT).
We finally highlight the potential astrophysical and cosmological relevance of combining
SZE and X-ray observations of radio-galaxy lobes.
For a direct comparison of our results with the X-ray observations of Erlund et al.
(2006) we use the same cosmological model with $H_0=71$ km s$^{-1}$ Mpc$^{-1}$,
$\Omega_0=1$ and $\Omega_{\Lambda} = 0.73$.

\section{IC emission from radio-galaxy lobes}

\begin{figure}
\begin{center}
 \epsfig{file=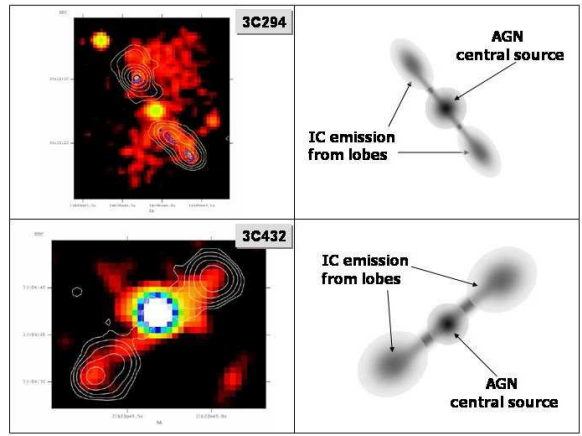,height=6.cm,width=8.cm,angle=0.0}
  \caption{The simplified geometry of the IC emission in the lobes of 3C432 and 3C294.
  The X-ray image of the sources have been taken from Erlund et al. (2006).
  The two lobes have an extension of  $L_{lobe} \approx 130$ kpc and a transverse size of
  $\approx 30-40$ kpc. The central AGNs and the kpc-scale lobe region are indicated by shaded
  circles and are not considered in this analysis. }
  \label{fig.jets}
\end{center}
\end{figure}

For the purposes of this paper we describe the system of the powerful radio-galaxy lobes
for which we calculate the SZE as a simplified geometry shown in Fig.\ref{fig.jets}.
In fact, the specific cases of the radio-galaxies 3C432 and  3C294 can be represented by
a central non-thermal AGN point-like source (we consider for simplicity that the
kpc-scale jet is within the same central AGN source) plus two symmetric lobes.
The linear size of the observed lobes of these sources is $\sim 130$ kpc (for both 3C294
and 3C432) and their transverse size is in the range $\sim 30$ to $\sim 40$ kpc.

The photon spectral index derived from the extended ICS X-ray emission is  $\Gamma \sim
1.57^{+0.27}_{-0.36}$ (for 3C432) and $\Gamma \sim 2.08^{+0.11}_{-0.08}$ (for 3C294). For
a power-law electron spectrum
 \be
 n_{e,rel}(\gamma) = n_{e0} \gamma^{-\alpha}
 \label{eq.espectrum}
 \ee
the index $\alpha = 2 \Gamma -1$ takes values $\alpha = 2.14$ (range $1.42 - 2.68$, for
3C432) and $\alpha=3.16$ (range $3-3.38$, for 3C294), respectively.\\
The energy density of the electronic population in the lobe is
 \be
 {\cal E}= \int_{\gamma_{min}}^{\gamma_{max}} d \gamma n_{e,rel}(\gamma) \gamma m_e c^2 \, .
 \label{eq.energy}
 \ee
For a value $\gamma_{min}=10^3$, $\gamma_{max}=10^5$ and $\Gamma = 1.7$ (i.e.,
$\alpha=2.4$), Erlund et al. (2006) found a total energy, $E_{tot} = \int dV_{lobe} {\cal
E}$ integrated over the volume $V_{lobe}$ of the radio lobe, of $E_{tot} \approx 3.7
\cdot 10^{59}$ erg (for 3C432) and $E_{tot} \approx 5.9 \cdot 10^{59}$ erg (for 3C294).
The energy density estimate however changes for different choices of $\gamma_{min}$ and
$\alpha$: while the spectral slope $\alpha$ can be stimated from X-ray data (see, .e.g.,
the results of Erlund et al. 2006), the parameter $\gamma_{min}$ is not constrained and
it is actually a free parameter in the range $\sim 1 - 10^2$ (see discussion above).\\
The previous parameters allow to calculate realistic estimates of both the X-ray ICS
emission from the radio-galaxy lobe and of the spectral and spatial features of the
associated SZE.

Inverse Compton (IC) scatterings of relativistic electrons on target cosmic microwave
background (CMB) photons give rise to a spectrum of photons stretching from the microwave
up to the gamma-ray band.
The IC power
\begin{equation}
P_{\rm IC}\left(E_\gamma,E\right) = c \, E_\gamma \int d\epsilon \, n(\epsilon) \,
\sigma(E_\gamma, \epsilon, E) \, .
 \label{eq:ICpower}
\end{equation}
is obtained by folding the differential number density of target photons $n(\epsilon)$
with the IC scattering cross section $\sigma(E_\gamma, \epsilon, E)$ given by the
Klein-Nishina formula:
\begin{equation}
\sigma(E_\gamma, \epsilon, E) = \frac{3 \sigma_T}{4 \epsilon \gamma^2}\,
G\left(q,\Gamma_e\right) \, ,
\end{equation}
where $\sigma_T$ is the Thomson cross section and
\begin{equation}
G\left(q,\Gamma_e\right) \equiv \left[  2 q \ln q + (1+2 q)(1-q) + \frac{\left(\Gamma_e q
\right)^2 (1-q)}{2 \left( 1+ \Gamma_e q\right)}\right]
\end{equation}
with
\begin{equation}
 \Gamma_e= 4 \epsilon \gamma / (m_e c^2) \;\; ; \;\;\;\;\;   q = E_\gamma /
\left[ \Gamma_e \left( \gamma m_e c^2 - E_\gamma \right)\right]\;.
\end{equation}
Here $\epsilon$ is the energy of the target photons and $E_\gamma$ the energy of the
up-scattered photon.
Folding the IC power in eq.(\ref{eq:ICpower}) with the the equilibrium spectrum of the
electrons, $n_{e,rel}$, the local emissivity of IC photons of energy $E_\gamma$ obtains:
\begin{equation}
 j_{\rm IC}\left(E_\gamma, r\right) = \int dE\, n_{e,rel}(E) P_{\rm IC}
 \left( E_\gamma,E \right)\;
 \label{eq:ICemiss}
\end{equation}
from which the integrated flux density spectrum obtains:
\begin{equation}
F_{\rm IC}(E_\gamma)= \int d V_{lobe} \, \frac{ j_{\rm IC}\left(E_\gamma, r\right)}{4
\pi\, D^2}\;,
 \label{eq:ICflux}
\end{equation}
or the IC brightness along the line of sight (los)
\begin{equation}
S_{\rm IC}(E_\gamma)= \int d \ell \, j_{\rm IC}\left(E_\gamma, r \right)\; .
 \label{eq:ICbrigtness}
\end{equation}
Here $D$ is the luminosity distance to the lobe source.
In Eq.~(\ref{eq:ICpower}) and Eq.~(\ref{eq:ICemiss}) the limits of integration over
$\epsilon$ and $E$
are set from the kinematics of the IC scattering which restricts $q$ in the range $1/(4
\gamma^2) \le q \le 1$.\\
The IC emissivity of a radio lobe (whose geometry is sketched in Fig.\ref{fig.jets}) in
the energy range $\Delta E_{\gamma}$ from $E_{\gamma,1}$ to $E_{\gamma,2}$ is given by
\be
 j_{IC} = {3 \over 4} A c \sigma_T {(k_B T_{CMB})^4 \over (h c)^3} {8 \pi \over \alpha
 +1}I(t)
 n_{e0} \bigg({\Delta E_{\gamma} \over A} \bigg)^{-(\alpha -1)/2} \; ,
 \label{eq:ICemiss_explicit}
\ee
where
 \be
 I(t) = \int d t t^3 (e^t -1)
 \label{It}
 \ee
with $t \equiv h \nu/k_B T_{CMB}$ and $E_{\gamma}=A (E/GeV)^2$ with $A= 2.13 \cdot
10^{-6}$ keV. The IC emissivity depends basically on the lower limit of integration
$E_{\gamma,1}$ for $\alpha > 2$, on the CMB temperature $T_{CMB}$ and on the electron
density normalization $n_{e0}$.

\section{The SZ effect in radio-galaxy lobes}

The SZE seen along the los to a radio-galaxy lobe is produced by the ICS of CMB photons
by the relativistic electrons confined into the radio-galaxy lobes. As such, the SZE is
-- in this case -- of a complete non-thermal nature, since the electron spectrum in the
lobe can be represented by a power-law distribution $n_{e,rel} \sim \gamma^{-\alpha}$
(see eq.\ref{eq.espectrum}).\\
The generalized expression for the SZE  which is valid in the Thomson limit ($\gamma h
\nu \ll m_e c^2$ in the electron rest frame) for a generic electron population in the
relativistic limit and includes also the effects of multiple scatterings and the
combination with other electron populations has been derived by Colafrancesco et al.
(2003) and we will refer to this paper for technical details.
According to these results, the spectral distortion of the CMB spectrum observable in the
direction of the radio-galaxy lobe is
 \begin{equation}
\Delta I_{\rm lobe}(x)=2\frac{(k_{\rm B} T_{CMB})^3}{(hc)^2}y_{\rm lobe} ~\tilde{g}(x) ~,
 \label{eq.deltai}
\end{equation}
with $x \equiv h \nu / k_{\rm B} T_{CMB}$, where the Comptonization parameter $y_{\rm
lobe}$ is
\begin{equation}
y_{\rm lobe}=\frac{\sigma_T}{m_{\rm e} c^2}\int P_{\rm e} d\ell ~,
 \label{eq.y}
\end{equation}
in terms of the pressure $P_{\rm e}$ contributed by the electronic population.
The spectral function $\tilde{g}(x)$ of the SZE is
\begin{equation}
 \label{gnontermesatta}
 \tilde{g}(x)=\frac{m_{\rm e} c^2}{\langle \varepsilon_{\rm e} \rangle} \left\{ \frac{1}{\tau_e} \left[\int_{-\infty}^{+\infty} i_0(xe^{-s}) P(s) ds-
i_0(x)\right] \right\}
\end{equation}
in terms of the photon redistribution function $P(s)$ and of  $i_0(x) = I_0(x)/[2 (k_{\rm
B} T_{CMB})^3 / (h c)^2] = x^3/(e^x -1)$, where
\begin{equation}
 \langle \varepsilon_{\rm e} \rangle  \equiv  \frac{\sigma_{\rm T}}{\tau}\int P_e d\ell
= \int_0^\infty dp f_{\rm e}(p) \frac{1}{3} p v(p) m_{\rm e} c
 \label{temp.media}
\end{equation}
is the average energy of the electron plasma (see Colafrancesco et al. 2003).
The optical depth of the electron population within the (cylindrical) lobe is
\begin{equation}
 \label{tau_p1}
 \tau_{\rm e}(p_1) = \sigma_T \int d \ell n_{\rm e,rel}(p_1)
\end{equation}
and takes the value
 \be
 \tau_{\rm e}(p_1) \approx 4.1 \times 10^{-8} \bigg[ {n_{\rm e,rel}(p_1) \over 10^{-6} {\rm cm}^{-3}}
 \bigg] \bigg({r_{\rm lobe} \over 10 {\rm kpc}} \bigg)
 \ee
along the los to the center of the lobe elongation axis.\\
For low values of $\tau_{\rm e}$, the following expression of $\Delta I_{\rm lobe}(x)$,
valid at first order in $\tau_{\rm e}$, holds:
 \be
\Delta I_{\rm lobe}(x) = 2\frac{(k_{\rm B} T_{CMB})^3}{(hc)^2}
 \tau_{\rm e} \bigg[ j_1(x) - j_0(x) \bigg] \; ,
 \label{eq.deltai_approx}
 \ee
where $j_1(x)= \int ds \, i_0(xe^{-s})P_1(s)$ and $j_0(x)=i_0(x)$.
The photon redistribution function $P_1(s)= \int dp f_{\rm e}(p) P_{\rm s}(s;p)$ with $s
= \ln(\nu'/\nu)$, in terms of the CMB photon frequency increase factor $\nu' / \nu = {4
\over 3} \gamma^2 - {1 \over 3}$, depends on the momentum ($p$ normalized to $m_ec$)
distribution, $f_{\rm e}(p)$ of the electrons which are filling the radio-galaxy lobe.\\
We describe here the relativistic electron plasma within the lobe with a single power-law
momentum spectrum
\begin{equation}
 \label{leggep1}
f_{\rm e,rel}(p;p_1,p_2,\alpha)=A(p_1,p_2,\alpha) p^{-\alpha} ~; \qquad p_1
\leq p \leq p_2
\end{equation}
where the normalization $ A(p_1,p_2,\alpha) = \frac{(\alpha-1)}
{p_1^{1-\alpha}-p_2^{1-\alpha}}$, with $\alpha =$ $2.14$ and $3.16$ for the cases of
3C432 and 3C294, respectively.\\
In the calculation of the SZE from the lobes, the relevant momentum is the minimum
momentum $p_1$ of the electron distribution (which is only marginally constrained by
X-ray observations) while the specific value of $p_2 \gg p_1$ is irrelevant for power-law
indices $\alpha > 2$, which are indicated by the electron spectra observed in radio
galaxy lobes we consider here.
In fact, for $\alpha > 2$ and $p_2 \gg p_1$, the normalization $A$ of the electron
spectrum $A \to {(\alpha -1) \over p_1^{1-\alpha}}$.\\
The value of $p_1$ sets the value of the electron density $n_{\rm e,rel}$ as well as the
value of the other relevant quantities which depend on it, namely the optical depth
$\tau_{\rm e}$, the pressure $P_{\rm e}$ and the energy density ${\cal E}_{\rm e}$ of the
non-thermal population.
The pressure $P_{\rm e}$, for the case of an electron distribution as in
eq.(\ref{leggep1}), writes as
\begin{eqnarray}
 \label{press_rel}
 P_{\rm e}&=&n_{\rm e,rel} \int_0^\infty dp f_e(p) \frac{1}{3} p v(p) m_e c \\
  & =& \frac{n_{\rm e,rel} m_e c^2 (\alpha
  -1)}{6[p^{1-\alpha}]_{\rm p_2}^{p_1}}
  \left[B_{\frac{1}{1+p^2}}\left(\frac{\alpha-2}{2},
   \frac{3-\alpha}{2}\right)\right]_{\rm p_2}^{p_1} \nonumber
\end{eqnarray}
(see, e.g., Ensslin \& Kaiser 2000, Colafrancesco et al. 2003), where $B_x(a,b)=\int_0^x
t^{a-1} (1-t)^{b-1} dt$.
The energy density for the same electron population writes as
\begin{eqnarray}
 \label{eneden_rel}
 {\cal E}_{\rm e} &=&n_{\rm e,rel} \int_0^\infty dp f_e(p) \bigg[ \bigg(1+p^2 \bigg)^{1/2} -1\bigg] m_e c^2 \\
  &=& \frac{n_{\rm e,rel} m_e c^2}{[p^{1-\alpha}]_{\rm p_2}^{p_1}}
  \bigg[{1 \over 2} B_{\frac{1}{1+p^2}} \left(\frac{\alpha-2}{2}, \frac{3-\alpha}{2}\right) \\
   & & + p^{1 -\alpha} \left( (1+p^2)^{1/2} -1 \right) \bigg]_{\rm p_2}^{p_1} \nonumber
\end{eqnarray}
and for a relativistic population of electrons ${\cal E}_{e} = 3 \cdot P_{e}$.
For an electron population with a double power-law (or more complex) spectrum, analogous
results can be obtained (see Colafrancesco et al. 2003 for details).
\begin{figure}
\begin{center}
 \epsfig{file=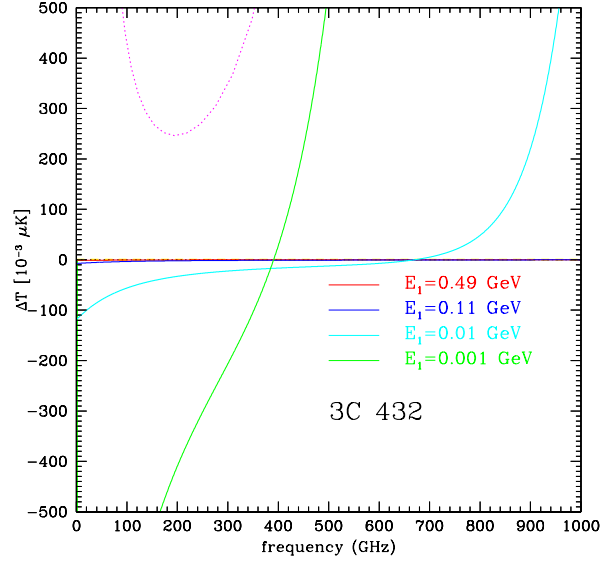,height=8.cm,width=8.cm,angle=0.0}
  \caption{The CMB temperature change $\Delta T$ due to the SZE from the lobes of
  3C432 is shown as a function of the frequency (in GHz).
  The spectral shape of $\Delta T$ is shown for different values of the minimum energy of the
  electrons: $E_{min}=0.49$ GeV (red), $E_{min}=0.11$ GeV (blue), $E_{min}=0.01$ GeV (cyan) and
  $E_{min}=0.001$ GeV (green). The level of brightness temperature due to CMB primary anisotropy
  (dashed yellow) and to the synchrotron emission of the radio-galaxy lobe
  (dotted magenta) are shown for comparison. The level of CMB brightness temperature due to the
  synchrotron emission in the lobe has been computed in the outer part of the radio lobe where it has a brightness of
  $1$ mJy/beam. Inner, higher radio brightness regions produce a CMB brightness temperature scaled
  up according to their relative intensity. }
  \label{fig.sz_3c432_dt}
\end{center}
\end{figure}
\begin{figure}
\begin{center}
 \epsfig{file=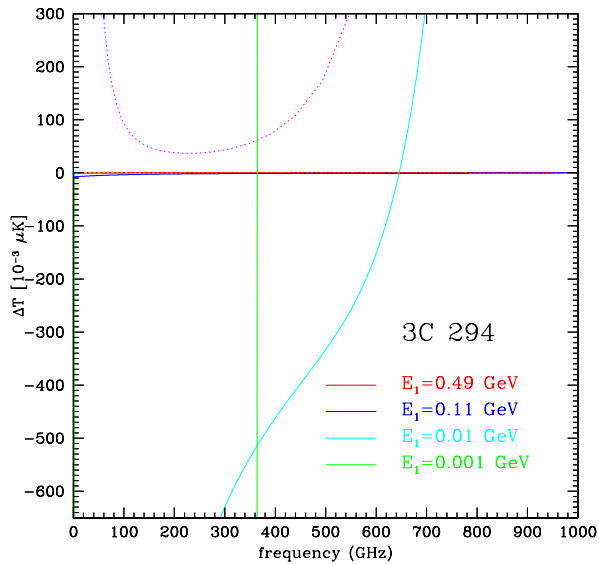,height=8.cm,width=8.cm,angle=0.0}
  \caption{Same as Fig.\ref{fig.sz_3c432_dt} but for the case of 3C294. The level of CMB brightness
  temperature due to the synchrotron
  radio emission in the lobe has been computed in the outer part of the radio lobe where it has a brightness of
  $2$ mJy/beam. Inner, higher radio brightness regions produce a CMB brightness temperature scaled
  up according to their relative intensity.}
  \label{fig.sz_3c294_dt}
\end{center}
\end{figure}

\noindent
The lobes of the radio galaxies here considered are assumed, for simplicity, to be
cylindrical with a radius $r_{\rm lobe} =20 kpc$. We also assume that the electron
spectrum given by eq.(\ref{leggep1}) does not depend on the position within the lobe.
The non-thermal electron spectrum is normalized to reproduce the total energy
$E_{tot}= \int dV_{lobe} {\cal E}_{e}$
of the electronic component in the lobe as derived from the Chandra X-ray observations,
after assuming (following Erlund et al. 2006) that: i) $\gamma_{min}=10^3$; and ii) the
values of $\alpha$ derived from the IC X-ray emission spectra observed for each source
here considered.\\
The total energy estimates given by Erlund et al. (2006) correspond, hence, to a
relativistic electron density $n_{\rm e,rel}(p_1) \approx 2.77 \times 10^{-9} {\rm
cm}^{-3}$ for a value $p_1 = 9.6\times 10^2$ and a slope $\alpha = 2.14$ for the case of
3C432, and to $n_{\rm e,rel}(p_1) \approx 7.5 \times 10^{-10} {\rm cm}^{-3}$ for the same
value $p_1 = 9.6\times 10^2$ and a slope $\alpha = 3.16$ for the case of 3C294.\\
Given the value of $E_{tot}$, the electron density $n_{\rm e,rel}$ decreases/increases
for increasing/decreasing values of $p_1$, according to eqs.(\ref{press_rel}) and
(\ref{eneden_rel}). Because the energy $E_{tot}(p_1=9.6 \cdot 10^2)$ for each source has
been obtained from the Chandra observation with a value of $E_{X,min}=2$ keV (see Erlund
et al. 2006), corresponding to a minimum energy in the electron spectrum of $E_{min}=
0.35 GeV (E_{X,min}/keV)^{1/2} \approx 0.49$ GeV, it should be definitely considered as a
lower limit to the actual energy/pressure within the lobe. As a consequence, the SZE
normalized to this energy value is also a lower limit to the actual SZE produced by the
non-thermal electron distribution within the lobe.
Thus, given the observed lobe pressure $P_{\rm e}$ and the electron spectrum slope
$\alpha$, the relative SZE takes different amplitudes and spectral features for the
values $n_{\rm e,rel}$ and $p_1$ which reproduce the pressure $P_{\rm e}(p_1)$.\\
We show in Figs.\ref{fig.sz_3c432_dt} and \ref{fig.sz_3c294_dt} the CMB temperature
change
 \be
 \bigg({\Delta T \over T}\bigg)_{lobe} = {{(e^x-1)^2}\over{x^4e^x}}
 \bigg({{\Delta I}\over I_0}\bigg)_{lobe}
 \label{eq.deltat}
 \ee
induced by the SZE in the case of the lobes of 3C432 and 3C294, respectively.
We stress that the lobe SZE is negative in sign at all microwave frequencies, up to the
crossover frequency, i.e. the frequency $x_{\rm 0}(P_{\rm e})$ at which the SZE from the
electron population with pressure $P_{\rm e}$ is zero (see Colafrancesco et al. 2003,
Colafrancesco 2005). Its amplitude increases for decreasing values of $p_1$, which
consistently yield increasing values of $n_{\rm e,rel}$, and is spatially located only in
the lobe regions (as also the IC X-ray emission).

\begin{figure}
\begin{center}
 \epsfig{file=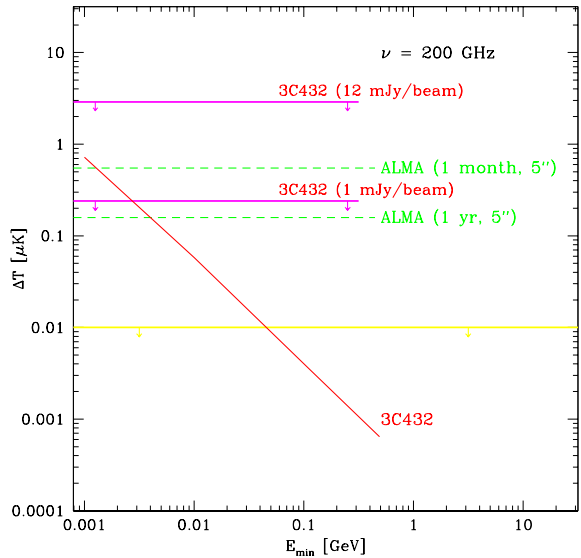,height=8.cm,width=8.cm,angle=0.0}
  \caption{The value of $\Delta T (\nu = 200 GHz)$ predicted for
  3C432 (red curve) as a function of $E_{min}$.
  Noise levels on $5$ arcsec beam produced by primary CMB anisotropies (yellow line)
  and synchrotron radio emission extrapolated to 200 GHz (using the low-frequency spectral slope
  for the source) from the lobes of 3C432 with brightness levels of 1 mJy/beam and 12 mJy/beam
  (thick magenta lines) are shown.
  The ALMA sensitivity for 1 month and 1 yr exposure with a $5$ arcsec beam
  are also shown for comparison.}
  \label{fig.DT_emin_3c432}
\end{center}
\end{figure}

\begin{figure}
\begin{center}
 \epsfig{file=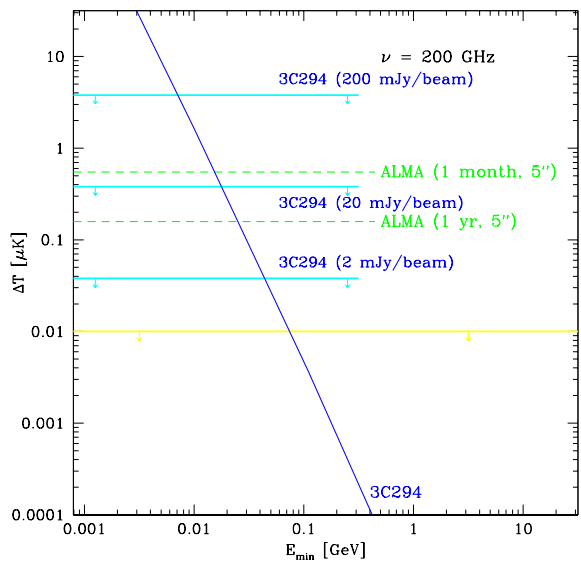,height=8.cm,width=8.cm,angle=0.0}
  \caption{The value of $\Delta T (\nu = 200 GHz)$ predicted for
  3C294 (blue curve) as a function of $E_{min}$.
  Noise levels on $5$ arcsec beam produced by primary CMB anisotropies (yellow line)
  and synchrotron radio emission extrapolated to 200 GHz (using the low-frequency spectral slope
  for the source) from the lobes of 3C294 with brightness levels of 2, 20 and 200 mJy/beam
  (thick cyan lines) are shown.
  The ALMA sensitivity for 1 month and 1 yr exposure with a $5$ arcsec beam
  are also shown for comparison.}
  \label{fig.DT_emin_3c294}
\end{center}
\end{figure}

The crossover frequency, $x_0(P_{e})$, of the SZE increases with increasing values of
$p_1$ (i.e. $E_{min}$) and, hence, of the lobe pressure/energy (see Figs
\ref{fig.sz_3c432_dt} and \ref{fig.sz_3c294_dt}). Thus, its determination can provide a
unique probe of the overall pressure/energy density of the electrons in the lobe and,
hence, of $E_{min}$ for a given electron spectrum (see Colafrancesco et al. 2003).
Figs. \ref{fig.DT_emin_3c432} and \ref{fig.DT_emin_3c294} show the dependence of the
amplitude of $\Delta T_{lobe}$ (evaluated at $\nu = 200$ GHz) associated to the lobes of
3C432 and 3C294 as a function of the lower cutoff $E_{min}$ of the electron spectrum.
Steeper power-law electron spectra produce larger SZE amplitudes at low values of
$E_{min}$ (for a fixed frequency) because they yield larger total energy/pressure and
optical depth for the electronic population and hence a larger amount of CMB photon
scattering off the non-thermal electrons residing in the lobes.

\section{Bias and confusion}
 \label{sect.bias}

Possible sources of confusion and contamination for the observation of the non-thermal
SZE in radio-galaxy lobes are the high-frequency tail of the synchrotron radio emission
from the lobes and CMB primary anisotropies on angular scales comparable with the radio
lobe size.

If the energy spectra of 3C294 and 3C432 extend down to $\sim 0.003$ GeV, or even lower
energies, the expected SZE at $200$ GHz from the lobes of these sources  are  $\sim 0.2$
$\mu$K for 3C432 and $\sim 30$ $\mu$K for 3C294 (see Figs. \ref{fig.DT_emin_3c432} and
\ref{fig.DT_emin_3c294}). These values correspond to diffuse fluxes of $\sim 0.16$
$\mu$Jy (for 3C432) and $\sim 24.3$ $\mu$Jy (for 3C294), respectively, observable with
ALMA.
To compare these values of the diffuse flux with those emerging from the lobes of 3C432
and 3C294, we considered here various levels of radio emission from these lobes (as
reported in Figs. 2 and 6 of Erlund et al. 2006) and we extrapolated their contribution
to 200 GHz, i.e. the frequency region where the contribution of the synchrotron spectra
is minimum (see Figs.\ref{fig.sz_3c432_dt} and \ref{fig.sz_3c294_dt}).\\
The level of synchrotron contamination depends in fact on the brightness level of the
radio lobes and it is more enhanced in the central part of the lobes rather than in their
external sides.

For 3C294 we found that the highest brightness level of 600 mJy/beam at 1.425 GHz, found
in the inner part of the lobes, with an energy spectral slope of $\alpha=2.14$
(corresponding to a synchrotron radio slope of $\alpha_r=(\alpha - 1)/2 =1.08$) produces
a brightness of $\approx$ 2.88 mJy/beam at 200 GHz. This yields a  CMB temperature
brightness of $\approx 11.5$ $\mu$K in a 5 arcsec beam.
Such high level of contamination renders extremely difficult to detect the SZE produced
in the lobes of 3C294 unless $E_{min}$ reaches very low values $\simlt 1$ MeV (see
Fig.\ref{fig.DT_emin_3c294}).\\
However, the absolute amplitude of the non-thermal SZE is higher than the synchrotron
brightness in the outer parts of the lobe, where the radio brightness drops down to
levels of $\approx 2$ mJy/beam (see Fig.\ref{fig.DT_emin_3c294}).
It is clear that such observational evidence implies (as already noticed by Hardcastle \&
Croston 2005) that either the magnetic field strength or the electron spectrum cannot
remain constant throughout the radio-galaxy lobes. The case in which the magnetic field
strength decreases throughout the lobe maximizes the chance of detection of the
associated SZE. However, also models in which both the magnetic field is somewhat weaker
and the electron spectrum is more depleted of high-energy electrons as a function of the
distance from the inner jet (a favoured model for the lobes of Pictor A, see Hardcastle
\& Croston 2005) tend to favour the detectability of the non-thermal SZE in the outer
lobe regions because such SZE effect is mainly sensitive to the low-energy part of the
electron spectrum which is likely not varying much with the distance from the inner jet.
Therefore, under the assumption that the electron spectrum does not depend on the
position within the lobe and the magnetic field strength decreases throughout the lobe,
the outer parts of the radio lobes of this source are the optimal regions to detect the
non-thermal SZE and therefore to set constraints to the value of $E_{min}$ and to the
overall energetics of the lobe.

For 3C432, the highest brightness level of 144 mJy/beam at 1.54 GHz, found in the central
part  of the radio lobes, with an energy spectral slope of $\alpha = 1.53$ (corresponding
to a synchrotron radio slope of $\alpha_r=(\alpha - 1)/2 =0.57$) produces a brightness of
$\approx$ 9 mJy/beam at 200 GHz. This yields a  CMB temperature brightness of $\approx
36$ $\mu$K in a 5 arcsec beam.
Such high level of contamination renders impossible to detect the SZE produced in the
lobes of 3C432 even for values of  $E_{min} \simlt 1$ MeV (see
Fig.\ref{fig.DT_emin_3c432}).
The absolute amplitude of the non-thermal SZE in the radio lobes of this source are
always lower than the synchrotron brightness in the outer parts of the lobe (even in the
optimal case in which the magnetic field strength decreases throughout the lobe), except
than at the outer boundaries of the radio lobes, where the radio brightness drops down to
levels of $\approx 1$ mJy/beam, providing a CMB temperature brightness of $\sim 0.25$
$\mu$K (see Fig.\ref{fig.DT_emin_3c432}). It would be, nonetheless, rather arduous to
detect this feeble signal due to both the low signal-to-noise ratio and to the long
($\sim 1$ year) exposure required for ALMA.

The different behaviour of the contamination of the two sources depends mainly on the
different synchrotron spectra here assumed, which render flat synchrotron spectra
extrapolated at microwave wavelengths more contaminating for the SZE measurements than
steep synchrotron spectra radio lobes. For the same reason, the lobes of 3C294, for which
a steeper electron spectrum is indicated by the IC X-ray data, produce a larger amount of
SZE signal for the case of low values of $E_{min}$ because most of the Comptonization of
the CMB radiation is produced by low-energy electrons.
It seems, therefore, that steep-spectra radio lobes are the best cases to detect the
non-thermal SZE expected in the extended lobes which are likely the sites of IC X-ray
emission.

In the previous estimates, we have assumed that the same electron spectrum producing the
X-ray IC emission is also holding for the more energetic electrons producing radio
synchrotron emission, and we have further calculated the contribution of the synchrotron
emission by extrapolating such a spectrum from low ($\sim 1.4-1.5$ GHz) radio frequencies
up to microwave ($\sim 200$ GHz) frequencies. Such a power-law extrapolation to 200 GHz
likely maximizes the level of synchrotron contamination to the SZE signal in the case in
which there is, instead, a break in the radio lobe spectrum at intermediate ($\sim 10 -
100$ GHz) frequencies.
Preliminary exploration at high frequencies (see e.g. Looney \& Hardcastle 2000;
Hardcastle \& Looney 2001) indicates that radio lobes can still emit at 100 GHz (in the
source rest frame, which is not really comparable to the 300 GHz source frame for the $z
\sim 2$ radio galaxies considered here). However, it can be argumented - on general
grounds - that, in the absence of an efficient and continuous particle (re-)acceleration
mechanism acting in the lobes, synchrotron and inverse-Compton losses will rapidly
deplete particles capable of radiating at these high frequencies. For equipartition field
strengths of $\sim$ a few $\mu$G, the break frequency due to loss processes is likely
found below $\sim 300$ GHz after $\simlt 2$ Myr from the electron injection event (see
discussion by Leahy 1991). Therefore, for electrons older than this age, the
extrapolation from low radio frequencies up to $\sim 200$ GHz will certainly overpredict
the level of synchrotron emission and of its contamination to the SZE signal.\\
Spatially resolved spectroscopic observations of radio lobes at microwave frequencies
($70-350$ GHz) will be definitely able to set constraints on the existence and frequency
location of the possible break of the radio synchrotron emission and, hence, on models of
the particle re-acceleration in these cosmic environments.
In the context of this explorative study aimed to detect the non-thermal SZE from the
radio lobes, it is instructive to notice that it is possible to set some constraints on
the slope and on the frequency of the synchrotron break from the requirement to have
little or no contamination of the SZE signal by radio emission from the lobe. Assuming,
for simplicity, that the synchrotron radio spectrum is given by a double power-law, $F =
F_0(\nu/\nu_0)^{\alpha_r}$ up to the break frequency $\nu_*$ and by a steeper spectrum $F
= F_*(\nu/\nu_*)^{\beta}$ at $\nu > \nu_*$, it is possible to show that the slope $\beta$
required in order to have a negligible contamination of the lobe synchrotron emission to
the SZE diffuse flux $F_{SZE}$ at frequency $\nu$ is
 \be
 \beta > \alpha_r + \bigg[ {log(F_0/F_{SZE})-\alpha_r log(\nu/\nu_0) \over log(\nu/\nu_*)}
 \bigg] \; ,
 \label{eq.beta}
 \ee
which cen be estimated in terms of measurable quantities $F_0, F_{SZE}, \alpha_r, \nu_0$
and $\nu_*$.\\
Nonetheless, it is clear that there will still likely be emission from regions with
ongoing particle acceleration (like, e.g.,  the inner jets and the hotspots of radio
galaxies) which might even substantially contaminate the SZE signal. These regions are,
however, spatially localized in the central parts of the extended lobes and certainly not
contaminating the outer regions of the lobes of a typical radio galaxy like 3C294, i.e.
the regions where there is the maximum probability to detect the non-thermal SZE signal.

Finally, it should be noticed that -- beyond the substantial contamination level of the
high-frequency tail of the synchrotron emission from the radio lobes -- a multi-frequency
analysis of both the synchrotron and SZE signals (which have quite different spectra in
the microwave band, see Figs. \ref{fig.sz_3c432_dt} and \ref{fig.sz_3c294_dt}) could
greatly help in subtracting the synchrotron signal from the SZE maps of radio lobes, even
in the case of strongly contaminated sources like 3C432.

The other source of contamination, i.e., the rms value of the primary CMB anisotropy
(predicted in the viable $\Lambda$CDM cosmological model) on a 5 arcsec angular scale is
of the order of $\sim 0.01$ $\mu$K and it is therefore a marginal contribution (at most)
to the SZE from the radio lobes in the two sources here considered. In addition,
contamination from CMB anisotropies to the SZE from the radio lobes would appear as a
constant level bias in Figs.\ref{fig.sz_3c294_dt} and \ref{fig.sz_3c432_dt}. Therefore, a
multifrequency analysis at microwave frequencies could clearly separate the CMB
anisotropy contamination (having a flat spectrum in the $\Delta T_{CMB} - \nu$ plot) from
the SZE (having a specific non-flat spectrum in the same plot).\\
We conclude that detection of the SZE from high-$z$ radio lobes is possible with the next
generation high-sensitivity, high-resolution experiments (like ALMA) if the electron
spectrum extends down to sufficiently low energies. Upper limits to the lobe SZE will
provide a low energy limit on the electron spectrum which is by far more constraining
than the limit set by the available X-ray observations. Similar observations can also be
made in nearby radio-galaxies with the SPT instrument that has a more moderate ($\sim 20$
arcsec - 1 arcmin resolution) but similar ($\sim \mu$K) sensitivity.

\section{Astrophysical and cosmological perspective}

The spatial observation of the SZE from radio-galaxy lobes has relevant astrophysical and
cosmological implications.

We stress that the constraints on $\gamma_{min}$ obtainable from SZE observations are
much more reliable than those obtained from X-ray IC emission, because the SZE depends on
the total pressure/energy density (that depends on the true $\gamma_{min}$) of the
electronic population along the los through the lobe, while the IC X-ray emission can
only provide an estimate of the electron energetics, and hence of $\gamma_{min}$, in the
high-energy part of the electron spectrum, namely at energies larger than $E_{min}
\approx 0.35 GeV (E_X/keV)^{1/2}$, where  $E_X$ is the lower energy threshold of the
X-ray detector (see discussion in Sect.1 above).
The conclusions of Blundell et al. (2006) on the discovery of a low energy cutoff in
powerful giant radio galaxies should, therefore, be rephrased in the previous framework.
These authors found, indeed, only that value of $E_{min} \sim 10^3$ consistent with the
lower energy threshold ($\sim 1$ keV) of their X-ray analysis of the radio-galaxy 4C39.24
(the previous value is of the same order of that found by Erlund et al. 2006, being in
fact obtained with the same X-ray instrument).
SZE observations of the lobes of this source should provide much more reliable estimates
of $\gamma_{min}$ and hence of the relative value of $\gamma_{min}$ in the compact radio
galaxy hotspots, where the energetic electrons have been likely originated.\\
As a consequence, also the scaling for the equipartition magnetic field, $B_{eq} \propto
\gamma_{min}^{-\delta}$ with $\delta = (2 \alpha_r -1)/(3+ \alpha_r)$ (see Blundell et
al. 2006) should be reconsidered: lower values of $\gamma_{min}$ consistent with the
total pressure/energy estimate in lobes, as likely provided by SZE observations, would
indicate higher values of $B_{eq}$ in the radio lobes.
Estimates of the overall B-field in the lobe by co-spatial IC X-ray and radio synchrotron
emission, $F_{radio}/F_{IC, Xray} \approx u_B / u_{CMB} \propto B^2$, strongly depend on
the assumption that the vast majority of X-ray emission is provided by ICS of CMB photons
(a similar assumption is usually done for estimates of wide-scale B-field in galaxy
clusters, see discussion by Colafrancesco et al. 2005). Partial contamination of the lobe
X-ray emission by other emission processes would, therefore, lower the value of $F_{IC,
Xray}$ and in turn increase the estimate of the overall B-field in the lobe.\\
In this context, it is interesting to notice that a better estimate of the overall
B-field in the lobe through this technique could be obtained by using the ratio
$F_{radio}/F_{SZE} \approx u_B / u_{CMB} $, where the SZE signal is a much safer
messenger of the CMB energy density which is reprocessed by all the possible Compton
scatterings induced by the full electron population in the radio-galaxy lobes.
In addition, the spatially resolved study of the SZE and of the synchrotron emission in
the lobes of radio-galaxies like 3C294 could provide interesting indication on the radial
behaviour of the magnetic field in these radio lobes.\\
A corollary of the previous considerations is that SZE signals are crucial to constrain
the lobe pressure and energetics, at least of the leptonic component, and to map the
leptonic pressure from the inner lobe regions out to lobe boundaries.\\
The study of the pressure evolution in radio lobes can also provide crucial indications
on the transition from radio lobe environments to the atmospheres of giant cavities
observed in galaxy cluster atmospheres (see, e.g., McNamara \& Nulsen 2007 for a review),
which seem naturally related to the penetration of radio-galaxy jets/lobes into the
intra-cluster medium (see, e.g., the extreme case of the giant cavities in the cluster
MS0735.6+7421). Previous studies of the spatial and spectral features of the SZE in giant
intracluster cavities (Colafrancesco 2005) will be naturally connected to the
developments of this preliminary study of the SZE for radio-galaxy lobes.
The development of experimental technology in microwave polarization will hopefully
provide the possibility to detect SZE polarization signals from radio-galaxy lobes and
allow the possibility to build a full 3-D tomography of radio lobes.

We finally stress that, under the hypothesis that the X-ray emission and the SZE from the
radio-galaxy lobes are produced by the same spatially-diffuse electronic population and
by the same mechanism (i.e. ICS of CMB photons), the X-ray and the SZE brightness have to
be co-spatial.
If the SZE along the los to the lobe is not found to be coincident with the relative IC
X-ray emission, then the nature of the lobe should be different from that of a diffuse
leptonic plasma, and its X-ray emission should be produced by a different mechanism.

To conclude this discussion, we outline here a few cosmological implications of the SZE
detectable from radio-galaxy lobes.
If the SZE and the IC X-ray flux of the lobes are produced by the same electronic
population, the ratio
\begin{eqnarray}
 {(\Delta T)_{lobe} \over F_{IC,lobe}} & = &
 {4 \over 3} {D^2 \over r_{lobe} L_{lobe}}
 \bigg[\tilde{g}(x) {(e^x -1)^2 \over x^4 e^x} \bigg]
 {m_e c^2 \over A} {1 \over c k_B } \nonumber \\
 & & \times {(hc)^3 \over (k_B T_{CMB})^3}
  {1 \over I(t)}
 {\alpha +1 \over \alpha -1}
 {\gamma_{min}^{-(\alpha-1)} \over ({\Delta E_{\gamma} \over A})^{-{(\alpha -1) \over 2}}
 }
 \label{eq.dtfluxratio}
 \end{eqnarray}
between the SZE temperature decrement $(\Delta T)_{lobe}$ and the X-ray IC flux
$F_{IC,lobe}$ integrated in the energy band $\Delta E_{\gamma}$ (see eqs.\ref{eq.deltai},
\ref{eq.y}, \ref{eq.deltat}, \ref{eq:ICflux} and \ref{eq:ICemiss_explicit}) provides an
estimate of $T_{CMB}$ at the radio-galaxy redshift in terms of measurable quantities and
fundamental constants. In other words, the ratio ${(\Delta T)_{lobe} \over F_{IC,lobe}}$
provides an operative way to test the evolution of the CMB temperature with redshift.
For a specific lobe source like 3C294, eq.(\ref{eq.dtfluxratio}) provides the scaling
 \begin{eqnarray}
 {\Delta T (100 GHz) \over L_{IC}(2-10 keV)} & \approx &
 0.72 \bigg({\mu K \over 10^{44} erg s^{-1}}\bigg)
 {({E_{min} \over 0.01 GeV})^{-(\alpha -1)} \over  ({E_{min} \over 0.01 GeV})^{-2.16}} \nonumber \\
 && \times {1 \over [2^{-(\alpha -1)/2} - 10^{-(\alpha -1)/2}]} \, .
 \label{eq.dtlicratio}
 \end{eqnarray}

\section{Concluding remarks}

We have shown in this paper that detailed predictions for the SZE from radio-galaxy lobes
can be obtained taking advantage of the constraints set on the energy/pressure by the
available X-ray observations.
The Chandra data for the lobes of 3C294 indicate that the associated SZE should be
visible by the next generation SZ experiments with $\sim$ arcsec and $\sim \mu$K
sensitivity, if its electron spectrum extends to quite low energies $E_{min} \simlt 1$
MeV.
Radio galaxy lobes with much flatter spectra, like those of 3C432, result instead swamped
by the large contamination of the synchrotron emission in the same radio lobes.\\
Steep spectra radio lobes whose IC emission is detected in the X-ray band by Chandra are
the best candidates to detect the associated SZE.

We have also shown that the spectral information on the SZE from lobes has a relevant
astrophysical impact.
The SZE provides, in fact, information on the energetics/pressure and on the optical
depth (geometry of the lobe): for the simple case of a power-law spectrum, it yields
information on the minimum energy of the electrons and on the minimum projected size of
the lobe. The case of more complex spectra can be treated in a similar way. This
information cannot be provided by the IC X-ray emission limited in a finite energy band (
e.g., $\sim 1-10$ keV), and hence to rather high values of $\gamma_{min} \sim 10^3$.\\
Moreover, since the zero $x_o(P_{e})$ of the SZE is moved to high frequencies depending
only on the total electron pressure/energy density (see Figs.\ref{fig.sz_3c432_dt} and
\ref{fig.sz_3c294_dt}), its detection provides direct information on the total
pressure/energy density of the electron population in the lobe.\\
We outline various potential uses of the non-thermal SZE detectable in radio lobes in the
astrophysical and cosmological contexts: i) constraints on the evolution of the
electronic pressure in the radio lobes and the transition to the larger-scale cavities of
the intra-cluster medium; ii) estimates of magnetic fields in lobes from a combination of
radio and SZE  measurements, possibly obtainable with single experiments with large
spectral coverage, like ALMA; iii) a full 3-D tomography of radio galaxy lobes obtainable
by means of the next generation microwave polarization experiments; iv) a test of the
evolution of the CMB temperature by using the ratio ${(\Delta T)_{lobe} \over
F_{IC,lobe}}$. These applications are still preliminary but also potentially important
with the advent of the next technology.

Most of the non-thermal SZE signal is predicted to be detectable in the outer regions of
the radio-galaxy lobes, where the radio synchrotron contamination dims out. Therefore,
SZE detectability is likely  possible with instruments with arcsec spatial resolution and
$\mu$K sensitivity (like ALMA) in distant sources, and with a somewhat lower resolution
instruments (like SPT) in nearby radio galaxies.\\
All the arguments presented in this paper show that the study of the SZE from
radio-galaxy lobes are highly complementary to the study of their IC X-ray emission and
are also crucial for their astrophysical and cosmological relevance.

\section*{Acknowledgments}
S.C. acknowledges fruitful suggestions made by the Referee of this paper, M.J.
Hardcastle, which have led to a substantial improvement of the presentation of the
results.

\bsp

\label{lastpage}

\end{document}